\documentclass[aip, reprint]{revtex4-1}

\usepackage[utf8]{inputenc}
\usepackage{amsmath, amsfonts, amssymb}
\usepackage{graphicx}
\usepackage{color, textcomp, graphics, graphicx, bm}
\usepackage{subfigure, array, xspace}
\usepackage[svgnames]{xcolor}

\usepackage[pdfstartview=FitV,bookmarks=true, linktocpage=true,colorlinks=true, urlcolor=blue,linkcolor=red, citecolor=blue]{hyperref}

\usepackage{fancyhdr}
\newcommand{\SiN}{Si$_3$N$_4$\@\xspace}
\newcommand{\supp}{Supplementary Information\@\xspace}

\begin{document}

\vspace*{\baselineskip}
\vspace*{\baselineskip}

\title{Imaging Gigahertz Zero-Group-Velocity Lamb Waves
\vspace*{\baselineskip}}

\author{Qingnan Xie}
\affiliation{School of Science, Nanjing University of Science and Technology, Nanjing 210094, People's Republic of China}
\author{Sylvain Mezil$^*$}
\author{Paul H. Otsuka}
\author{Motonobu Tomoda}
\affiliation{Division of Applied Physics, Graduate School of Engineering, Hokkaido University, Sapporo 060-8628, Japan}
\author{J\'er\^ome Laurent}
\affiliation{Institut Langevin, ESPCI ParisTech, CNRS, 1 rue Jussieu, 75238 Paris Cedex 05, France}
\author{Osamu Matsuda}
\affiliation{Division of Applied Physics, Graduate School of Engineering, Hokkaido University, Sapporo 060-8628, Japan}
\author{Zhonghua Shen}
\affiliation{School of Science, Nanjing University of Science and Technology, Nanjing 210094, People's Republic of China}
\author{Oliver B. Wright}
\affiliation{Division of Applied Physics, Graduate School of Engineering, Hokkaido University, Sapporo 060-8628, Japan}

\begin{abstract}
We image GHz zero-group-velocity (ZGV) Lamb waves in the time domain by means of an ultrafast optical technique, revealing their stationary nature and their acoustic energy localization in two dimensions. 
The acoustic field is imaged to micron resolution on a nanoscale bilayer consisting of a silicon-nitride plate coated with a titanium film. Temporal and spatiotemporal Fourier transforms combined with a technique involving the intensity modulation of the optical pump and probe beams gives access to arbitrary acoustic frequencies, allowing ZGV modes to be isolated. The dispersion curves of the bilayer system are extracted together with the Q factor and lifetime of the first ZGV mode. Applications include the testing of bonded nanostructures. 
\end{abstract}

\pacs{}
\keywords{ultrasonics, ultrafast, zero group velocity, acoustics, nanostructure}

\date{\today}

\maketitle

 \section*{}
 \newpage
 \section*{}
 \newpage 


Waveguides channel propagating waves with reduced losses thanks to the confined dimensions, and are widely used in optics and acoustics. They are dispersive, i.e., the phase  and the group velocities differ. Zero-group-velocity (ZGV) modes are particular points in a dispersion relation where the group velocity vanishes whereas the phase velocity remains finite. These modes can be found in most waveguide geometries (e.g., fibres and cylinders~\cite{ranka00, laurent15}, plates~\cite{holland03, prada05}, etc.), and have the advantage of combining energy localization and high Q factor. Many applications take advantage of these unique properties. In optics, they are, for example, implicated in soliton propagation,  pulse compression and microcavity confinement~\cite{ranka00, milian14, ibanescu05}. In acoustics, they are important in structural testing using Lamb waves, i.e., in geometries with free-surface boundary conditions and for which the acoustic wavelength is of the same order as the thickness~\cite{rayleigh89}.  Acoustic ZGV Lamb modes offer, for instance, methods for estimating the Poisson's ratio~\cite{clorennec07}, thin layer thicknesses~\cite{ces11b}, elastic constants~\cite{ces12, grunsteidl16}, and interfacial stiffnesses between bonded plates~\cite{mezil14, mezil15b} or fatigue damage~\cite{yan18}. ZGV Lamb modes have their energy trapped within a specific lateral region, offering a local measurement. These modes can be accessed by contactless excitation and detection, achievable with air-coupled transducers~\cite{holland03}, electromagnetic acoustic transducers~\cite{dixon01} or lasers~\cite{prada05, clorennec07, ces11b, ces12, grunsteidl16, mezil14, mezil15b, yan18}, often working in the kHz-MHz range. 

The observation of ZGV Lamb modes in plates has been extended up to the GHz range by the use of interdigital transducers~\cite{yantchev11} or intensity-modulated continuous lasers~\cite{balogun07}, but only up to $\sim$40~MHz by the use of pulsed lasers~\cite{yan18}. Observations of GHz ZGV Lamb waves with pulsed lasers has, however, not proved possible owing to the extremely sharp resonances associated with ZGV modes\textemdash exhibiting Q factors up to 14700~\cite{clorennec06}, for example\textemdash whereas acoustic frequencies are usually limited to integral multiples of the laser repetition rate. Ultrashort-pulse lasers are ideal for time domain imaging, but to our knowledge the imaging of ZGV modes in two dimensions has not been investigated.

In this paper, we image a GHz ZGV Lamb mode in a nanoscale bilayer consisting of a silicon nitride plate coated with polycrystalline titanium by means of a time-resolved two-dimensional (2D) imaging technique incorporating an ultrashort-pulse laser\cite{tomoda14}. We overcome the above-mentioned frequency limitation by the use of arbitrary-frequency control that takes advantage of the sidebands introduced by additional intensity modulation in the laser beam paths~\cite{kaneko14, matsuda15}. In particular, we identify and isolate the first ZGV mode, its associated Q factor and its lifetime. The experimental dispersion curves of the bilayer are obtained, clearly showing the location of the ZGV mode in frequency-wavevector space and its acoustic energy localization.

\section*{Results}
\begin{figure*}[t]
\centering
\includegraphics[width=\textwidth]{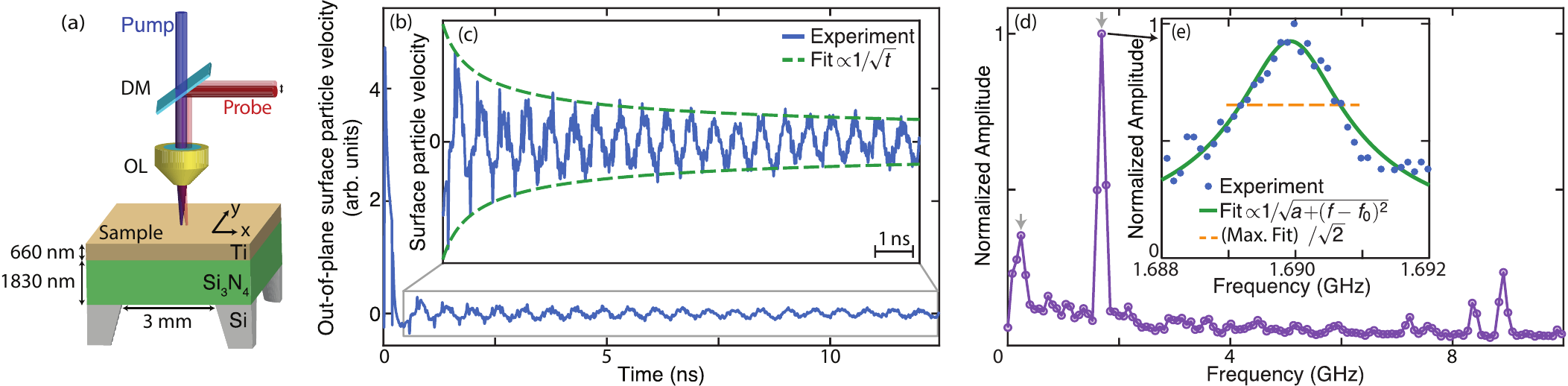}
\subfigure{\label{fig:sample}}
\subfigure{\label{fig:signaltime}}
\subfigure{\label{fig:timezoom}}
\subfigure{\label{fig:fouriersignal}}
\subfigure{\label{fig:evolpeak}}
\caption{\textbf{Sample, surface particle velocity variation and acoustic frequency spectrum.}  (a)~Schematic diagram of the sample. DM: dichroic mirror, OL: objective lens. (b)~Out-of-plane surface particle velocity measured for co-focused pump and probe beams. (c)~Zoom-in of the ZGV Lamb mode amplitude temporal evolution and fit $\propto 1/\sqrt{t}$. (d)~Frequency spectrum obtained after a temporal Fourier transform.  Arrows indicate the two modes imaged in Fig.~\ref{fig:evol2D}. (e)~Evolution of the normalized amplitude in the frequency window $1.688\leqslant f \leqslant 1.692$~GHz. The solid line is a fit to the experimental data in the form of the square root of a Lorentzian. The dashed line represents $1/\sqrt{2}$ (i.e., -3~dB) of the maximum amplitude of the fit.}
\end{figure*}

\vspace{-0.5cm}
%
%
\noindent\textbf{Experimental setup and theoretical model ---}  The sample, depicted in Fig.~\ref{fig:sample}, consists of a square silicon nitride plate (of approximate composition \SiN) of thickness 1830~nm coated with a 660~nm sputtered polycrystalline titanium film.  Experiments were carried out with an optical pump-and-probe technique combined with a common-path Sagnac interferometer, that allows the possibility of imaging \cite{tachizaki06}. A pump optical beam focused to micron-sized region of the sample surface ($x$, $y$) is used to excite Lamb waves, and the temporal or spatiotemporal evolution of the normal surface particle velocity is obtained with $\sim$100 ps time-resolution with a micron-sized probe beam spot and by varying the pump-probe delay time with a delay line. Deformations occur throughout the sample thickness (as expected for Lamb waves). For the initial experiments the pump and probe beams are co-focused to one point on the sample, and for the later experiments the probe beam is scanned over the surface in 1D or 2D. The arbitrary-frequency technique involving the modulation of the pump beam and/or probe beam is described in detail in Methods.

In order to facilitate the identification of a ZGV mode, we calculate the acoustic dispersion relation of our bilayer structure, making use of literature elastic constants~\cite{briggs92livre} as well as layer thicknesses obtained from ultrafast pulse-echo measurement (see \supp for details). The first ZGV Lamb mode is predicted to be at frequency $f_1^{\rm{th}}=1.7248$~GHz and wavenumber $k_1^{\rm{th}}=0.620$~\textmu m$^{-1}$. Two other ZGV Lamb modes below 10~GHz are predicted near 3 and 7~GHz (see Table~I in the \supp).  We concentrate in this paper on the first ZGV Lamb mode, which has a significant out-of-plane acoustic displacement component. 
\\

 \noindent{\textbf{Detection of a GHz ZGV Lamb mode ---}}
  Figure~\ref{fig:signaltime} shows the temporal evolution of the out-of-plane surface particle velocity for an acoustic frequency of $f=1.6900$~GHz for co-focused pump and probe spots of a few microns in width on the sample (see Methods).  This frequency was isolated initially by the help of the above theoretical prediction for the first ZGV resonance and subsequently by frequency tuning.  A thermal background variation is subtracted using a polynomial function. This clearly reveals the ZGV resonance, a mode with a lifetime 
  greater than the laser repetition period of $\sim$12 ns (as explained in detail later). An enlarged view of the data is shown in Fig.~\ref{fig:timezoom}. Two effects contribute to the decay in amplitude. The first originates from the second-order term in the dispersion relation $\omega(k)$ in the vicinity of the ZGV resonance, where $\omega$ is the angular frequency and $k$ the acoustic wavenumber. This produces a $1/\sqrt{t}$ decrease with time $t$, as previously observed at lower frequencies~\cite{prada08b}. The second originates from viscoelastic attenuation, yielding a decay $\propto e^{-t/\tau_0}$, where $\tau_0$ is the mode lifetime, and can be observed only after a certain time has passed. 
  The amplitude decrease in the present case can be fitted to a good approximation with a $1/\sqrt{t}$ function (using least-squares, as with all subsequent fits), as shown in Fig.~\ref{fig:timezoom}. For our time window equal to $\sim$20 periods of the ZGV mode, no exponential decrease is visible, precluding a derivation of $\tau_0$ by this method. 
  
 The experimental spectrum obtained from the combination of the different scanned frequencies is displayed in Fig.~\ref{fig:fouriersignal} (see Methods for details). Besides the strong ZGV resonance at 1.6900~GHz, other peaks are observed at 0.2391, 7.2365, 8.3594 and 8.9202~GHz, which are selected a) owing to the choice of pump modulation frequencies, as shown in Methods, b) because with co-focused optical pump and probe spots one only detects modes that do not leave the excitation region, i.e. modes with zero or low group velocity, and c) because with the probe-laser normal incidence only modes with significant normal displacement are detected.
The detected modes lie on the $qA_0$, $qA_4$, $qS_8$ and $qS_5$ branches of the acoustic dispersion relation, where $qA$ refers to quasi-antisymmetric and $qS$ refers to quasi-symmetric Lamb waves (see Supplementary Information). The observed ZGV mode thus corresponds to the $qS_1$ ZGV. 

\begin{figure*}[t]
\centering
\includegraphics[width=\textwidth]{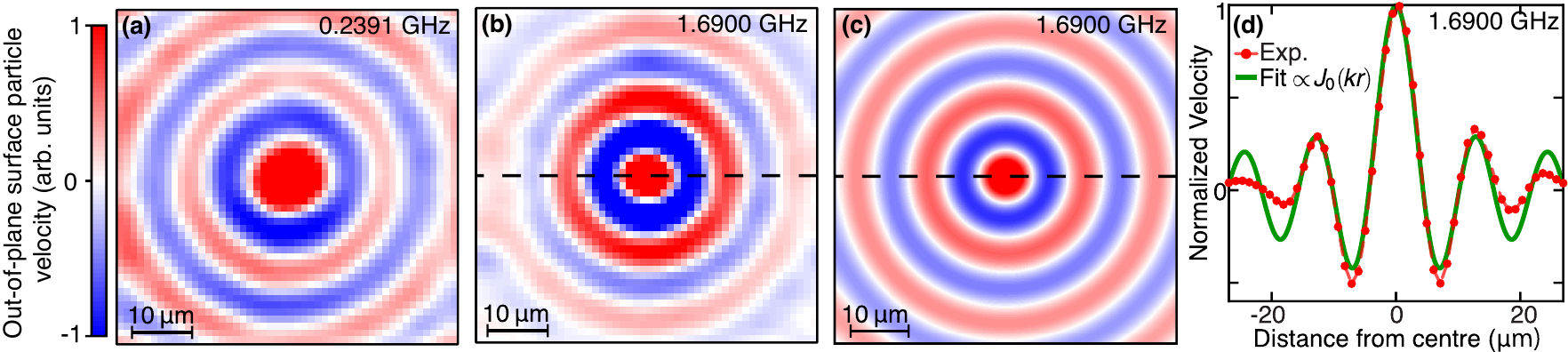}
\subfigure{\label{fig:animation02391}}
\subfigure{\label{fig:animation16900}}
\subfigure{\label{fig:fit2D}}
\subfigure{\label{fig:fit1D}}
\caption{\textbf{Imaging a ZGV Lamb mode.} (a--b)~Normalized images of the measured out-of-plane surface particle  velocity at (a)~0.2391 and (b)~1.6900~GHz~GHz. The $x$ and $y$ axis directions are shown in Fig.~\ref{fig:sample}. Animations are viewable in the Supplementary Information. (c)~2D acoustic field  based on a fit to a Bessel function $J_0(kr)$. (d)~Normalized out-of-plane surface particle velocity evolution of the experimental (dots) and fitted (solid line) data as a function of radial distance (dashed lines in (b)--(c)).} 
\label{fig:evol2D}
\end{figure*}

The experimental ($f_1^{\rm{exp}}=1.6900$~GHz) and theoretical ($f_1^{\rm{th}}=1.7248$~GHz) ZGV frequencies are in reasonable agreement. The residual $\sim$2~\% mismatch may be caused either by a difference in the \SiN layer parameters (see \supp) or to imperfect adhesion between the two layers~\cite{mezil14,mezil15b}. The two other ZGV Lamb modes, predicted near 3 and 7~GHz, could not be detected. This is thought to be owing to a mismatch between the probed frequencies and the ZGV frequencies (see Methods) as well as, for the 7~GHz peak, to the smaller out-of-plane surface displacement expected for this mode (see Supplementary Information).

Figure~\ref{fig:evolpeak} shows the amplitude-frequency relation around $f_{1}^{\rm{exp}}$. In order to extract the associated Q factor, we fit with the square root of a Lorentzian function, $b/\sqrt{a+(f-f_0)^2}$, where $a, b$ and $f_0$ are fitting parameters, yielding $f_0=1.6899$~GHz and $Q=1150$. This of the same order of magnitude as the Q factors previously reported for thin composite silicon-nitride/metal/oxide plates suspended by thin beams and driven in thickness longitudinal resonances at similar frequencies \cite{pang2004high}. For ZGV modes, Q factors up to 14700 at MHz frequencies have been observed in other materials \cite{clorennec06,prada09}, values which depend strongly on the ultrasonic attenuation at the frequency in question, as discussed in the next section, as well as parameters such as the possible imperfect adhesion between the two layers.
\\
 
%
%
\noindent{\textbf{Imaging a ZGV Lamb mode ---}}
\noindent The spatiotemporal evolution of the acoustic field is imaged by scanning the probe spot in 2D over a $55\times55$~\textmu m$^2$ area using 301 frames.  We specifically target the ZGV mode (at $f_1^{\rm{exp}}$) and the branch $qA_0$ at $\sim$0.24~GHz, indicated by the two downward pointing arrows in Fig.~\ref{fig:fouriersignal}.  To eliminate unwanted frequency components, filtering of small and high wavenumbers is conducted in Fourier space. We thereby extract the amplitude field, as shown in Fig.~\ref{fig:animation02391} and \subref{fig:animation16900} for $f=0.2391$ and $f_1^{\rm{exp}}=1.6900$~GHz, respectively (also viewable as animations in a Multimedia view). The 1.6900 GHz data represents, to our knowledge,  the first experimental 2D movie of a ZGV mode, extending the 1D observations of Laurent {\it{et al.}}~\cite{laurent14}. Comparing the animations, one can immediately ascertain that the ZGV mode is not propagating, in striking contrast to the $qA_0$ branch, which corresponds to propagating modes. 

For a single-frequency point-excited wave in two dimensions, the radial form of the out-of-plane displacement is proportional to $J_0(kr)$, where $J_0$ is the first-order Bessel function and $r$ is the radial distance (see, e.g., Ref.~\citenum{wright02}).  This has previously been confirmed by theory and by simulation for ZGV Lamb modes~\cite{prada08b,balogun07}. Figure~\ref{fig:fit2D} shows a fit to our data at 1.6900~GHz with the function $J_0(k r$), which gives reasonable agreement, also clear from the cross section shown in Fig.~\ref{fig:fit1D}. The value of $k$ for the best fit,  $k_{1}^{\rm{exp}}=0.55\pm0.13$~\textmu m$^{-1}$, is in fair correspondence with the predicted value $k_{1}^{\rm{th}}=0.620$~\textmu m$^{-1}$ from the theoretical dispersion relation \footnote{The fit does not take into account the smoothing due to the finite spot diameters, but we verified that its effect has a negligible influence on the fitted curve.}. One can notice that the experimental amplitude on the first main side lobes is slightly larger than the theoretical Bessel fitted function, which is in agreement with the results obtained by Laurent \textit{et al.}\cite{laurent14}. Away from the central peak, the experimental amplitude is lower; these effects can be attributed to high-frequency material losses.

Our measurement of Q factor ($Q=1150$) and wavenumber  ($k_{1}^{\rm{exp}}=0.55$~\textmu m$^{-1}$) for the ZGV mode at 1.6900~GHz allows an estimate of the lifetime $\tau_0$. As a first step, for a free-standing layer of a single isotropic material and considering only viscoelastic losses \cite{prada08b, clorennec06}, the spatial attenuation coefficient is given by
\begin{equation}
\alpha=\frac{k}{2Q},
\label{eq:attenuation}
\end{equation}
where $\alpha$ is an effective attenuation coefficient (in m$^{-1}$). This relation yields $\alpha=240$~m$^{-1}$ for the above-mentioned ZGV mode. One can then make use of the relation $\tau_0$=$1/(\alpha v_p)$, where $v_p$ is the phase velocity to find $\tau_0$$\approx$0.22~\textmu s. This is much longer than the time $\sim$10~ns, for an acoustic mode with a typical sound velocity $\sim$5~km/s to leave the imaged region of $55\times55$~\textmu m$^2$ in which the ZGV amplitude is significant.

The value of $\alpha$ obtained can be compared with other high frequency measurements. Assuming an $f^2$ variation in $\alpha$, as expected from viscous losses, our value $\alpha=240$~m$^{-1}$ lies between those extrapolated for longitudinal waves in silicon nitride (29 m$^{-1}$) or polycrystalline titanium (2300~m$^{-1}$) to a frequency of 1.69 GHz~\cite{mansfeld2001acoustic, emery2006acoustic}.
 A detailed comparison is difficult because ZGV modes couple both longitudinal and transverse strain  components, which are distributed over the two layers (see Fig.~S2 in the Supplementary Information).\\

\begin{figure}[t]
\centering
\includegraphics[scale=1]{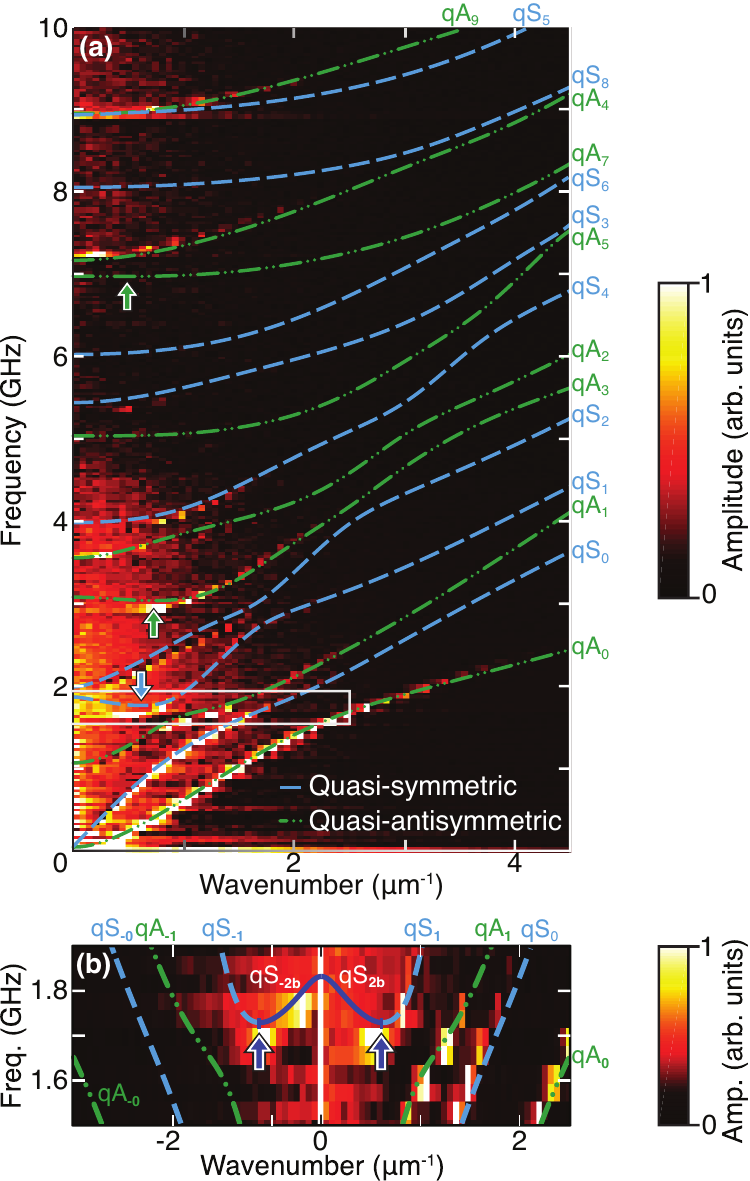}
\subfigure{\label{fig:dispcurves}}
\subfigure{\label{fig:zoomdispcurves}}
\caption{\textbf{Dispersion curves of the  bilayer.} (a--b)~Experimental (image) and theoretical (dashed and dotted-dashed lines) dispersion curves for (a)~$0\leqslant f \leqslant 10$~GHz and $0\leqslant k \leqslant 4.5$~\textmu m$^{-1}$ (b)~at frequencies near the first ZGV mode for $1.5\leqslant f \leqslant 1.9$~GHz and $-2.5\leqslant k\leqslant 2.5$~\textmu m$^{-1}$. The solid navy blue section of the dispersion curves corresponds to the portions where the phase and group velocities are anti-parallel. 
(a)~The box corresponds to the enlarged region in (b) for $k>0$. (a-b)~Arrows indicate ZGV points. Plots are independently normalized. }
\label{fig:alldispcurves}
\end{figure}

%
%
\noindent{\textbf{Dispersion curves --- }} 
\noindent More complete information is available with an experimental knowledge of the Lamb-wave dispersion curves. To this end, the pump beam is focused to a micron-sized line source with a cylindrical lens (see Methods). The probe beam spot is scanned along the direction $x$ ($>0$, see Fig.~\ref{fig:sample}) perpendicular to this line source over distances up to 100~\textmu m in 0.05~\textmu m steps to obtain the spatiotemporal variation of the acoustic field. 
 A 2D FT (temporal FT and 1D spatial FT) allows the extraction of the acoustic dispersion curves, as shown in Fig.~\ref{fig:dispcurves} for both experiment and theory, which show good agreement for the modes visible in experiment (bright regions). The positive wavenumbers in Fig.~\ref{fig:alldispcurves} correspond to waves with phase velocity along the $+x$ direction.  The three branches $qA_0$, $qS_0$ and $qA_1$, observed here below $\sim$2~GHz, are detected for a wide range of wavenumbers, unlike branches with higher frequencies, which are detected only for small wavenumbers $k$. 
The first ZGV mode is clearly observed at $k=0.54\pm$0.03~\textmu m$^{-1}$, in agreement with the value extracted in the 2D-scan experiment. The second ZGV mode, predicted at $f_{2}^{\rm{th}}=3.0024$~GHz and $k_{2}^{\rm{th}}=0.732$~\textmu m$^{-1}$ is evident at $f_{2}^{\rm{exp}}=2.972$~GHz and $k_{2}^{\rm{exp}}=0.83\pm0.03$~\textmu m$^{-1}$, in fair agreement. This mode may not have been observed in the previous experiment with co-focused laser beams owing to the different excited wave vector spectrum ($k_1^{th}<k_2^{th}$).
Other peaks that were previously detected at 0.2391, 7.2365 and 8.9202~GHz are again evident on the branches $qA_0$, $qA_4$ and $qS_5$, respectively. (The mode previously detected at 8.3594~MHz is absent here, and no corresponding mode appears on the dispersion relation. Its previous appearance in the spectrum of Fig.~\ref{fig:fouriersignal} seems likely to be an experimental artifact.) Some branches, such as $qA_2$ near $3.6$~GHz as well as $qS_0$ and $qA_1$, are revealed only in this new measurement. We attribute residual discrepancies between the theoretical and experimental dispersion curves to uncertainties in the elastic parameters and in the layer thicknesses, and to the possible imperfect layer bonding~\cite{vlasie03}. 

The dispersion curves in Fig.~\ref{fig:zoomdispcurves} are a magnified view of Fig.~\ref{fig:dispcurves}, this time including negative wavenumbers corresponding to $-x$-directed propagation. The main features are as follows:
  \begin{enumerate}
  \item[1)] for the $qA_0$, $qS_0$ and $qA_1$ branches and the region $(\omega,k)$ with $|k|$ larger than that for the $qS_1$ ZGV (the latter indicated by a downward-pointing arrow in Fig.~\ref{fig:dispcurves}), the intensity is much greater for $k>0$;
  \item[2)] for the region $(\omega,k)$ with $|k|$ smaller than that for the $qS_1$ ZGV, the intensity is much greater for $k<0$;
  \item[3)] for the region $(\omega,k)$ in the vicinity of the $qS_1$ ZGV, near wavenumber $k_1$ and frequency $f_1$, the intensities for $k>0$ and $k<0$ are comparable.
  \end{enumerate}
	
\section*{Discussion}
\vspace{-5mm}
\noindent Because of the symmetry of the excitation with respect to coordinate $x$, acoustic waves with positive and negative $k$ should be generated with equal amplitude and initially form a single wave packet located at the excitation point. This wave packet is eventually broadened and split because of the acoustic dispersion and the broad distribution of positive and negative values of $k$. Its wave components are plane-wave modes spreading over all 2D space, but in general, their sum is observable as a wave packet only if their phase constructively interferes.  In other words, there are two possibilities for observing no acoustic field: a complete lack of such modes or the overlapping of several modes with random phase.  (Compare the Fourier transform of a $\delta$-function: it contains an infinite number of fully spreading plane waves, but is only finite at a point in space.) Since we observe the waves in the region $x>0$, the waves we can primarily observe are restricted to those with $v_g\geqslant 0$.  Waves with $v_g<0$ exist, but give negligible contribution to the waves we observe because we only probe for positive $x$ values over a region with little overlap with the pump spot. 

The feature 1) above can therefore be attributed to the waves having co-directed group  velocity $v_g$ and phase velocity $v_p$.  With reference to the dispersion curves, all $k$ values involved in feature 1) show positive group velocity $v_g>0$.  On the other hand, feature 2) can be attributed to the waves having anti-parallel $v_g$ and $v_p$; the $k$ values involved in feature 2) can be seen to be negative, but are associated with a positive slope on the $\omega$-$k$ plot, i.e., they possess $v_g>0$ but with $v_p<0$. The feature 3) can be attributed to the waves with zero or very small group velocity.  The wave packet consisting of these near-zero or zero-$v_g$ wave components remains almost stationary, and both positive and negative $k$ components contribute to this wave packet over an extended period of time. These features are consistent with those of Philippe \textit{et al.}~\cite{philippe2015focusing}. In Fig.~\ref{fig:zoomdispcurves} we label the part of the +$k$ $qS_{1}$ branch with the negative slope as $qS_{2b}$ (and $qS_{-2b}$ for the equivalent $-k$ part), in agreement with standard nomenclature~\cite{meitzler65, prada08}, where $b$ stands for `backward wave'. 
All other branches with finite positive group velocity are observed only for $k>0$, with the exception of $qS_5$ near 9~GHz, for which $v_g$ is small for low $k$, and this branch is observed for both positive and negative $k$ with similar amplitude. Only the branch containing the ZGV point at 1.6900~GHz has a greater amplitude for negative $k$ compared with that for positive $k$.  

In conclusion, we image a zero-group-velocity Lamb mode in two dimensions. We apply an ultrafast time-domain technique with arbitrary GHz-acoustic frequency control to a nanoscale bilayer consisting of a silicon-nitride plate coated with a titanium film to provide this observation at unprecedented  frequencies in the GHz range. Our combination of both time-domain 1D and 2D optical scanning methods provides a comprehensive probe for the ZGV dynamics. We isolate the $qS$$_1$ 
 ZGV mode at $\sim$1.7 GHz and probe its spatial and temporal characteristics, including its Q factor $\sim$1000. The spatial form of this quasi-point-excited ZGV is directly verified by Fourier analysis to correspond closely to the expected Bessel function in 2D, allowing its wavenumber to be derived and its ZGV nature to be directly verified in the time domain from its relatively long lifetime $\sim$0.2~\textmu s. Experimental dispersion curves of this bilayer system are also obtained, and show good agreement with a theoretical model and reveal other ZGV modes. 

Applications of real-time imaging of ZGV Lamb modes include detecting defects in adhesion and deviations in interfacial stiffnesses, so this high frequency imaging technique should provide new avenues for evaluating and quantifying the mechanical integrity of nanostructures.

%
%
\section*{Methods}
\vspace{-5mm}
\noindent\textbf{Measurement setup ---} 
A Ti:sapphire pulsed laser produces a series of $\sim$100~fs width  pulses at repetition rate $f_\text{rep}= 80.38$~MHz and wavelength 830~nm. Part of this beam is frequency doubled ($\lambda=415$~nm) and used for the pump with a pulse energy of $\sim$0.15~nJ. It is intensity modulated with an acousto-optic modulator (AOM) at frequency $f_p$. For experiments with a circular pump spot, the pump beam is focused on the Ti film side of the sample at normal incidence with a $\times 50$ objective lens  to a 4.2~\textmu m radius (at $1/e^2$ intensity) to optimize the first ZGV Lamb mode generation (see \supp).  This generates, in the centre of the pump spot, an instantaneous temperature rise of $\sim$80~K after each laser pulse and a steady state temperature rise of $\sim$50~K~\footnote{The steady state temperature rise $T$ of the sample at the centre of the optical pump spot is estimated by considering a finite-sized effectively 2D circular plate and approximating the laser intensity profile to a top hat distribution. Under such assumptions, the solution of the heat diffusion equation gives $T=P/(2\pi t \kappa)\times[\text{ln}(a/w)+1/2]$, where $P$ is the power absorbed by the sample ($P=P_0 T_0 (1-R_0) $ with $P_0=6$~mW the measured incident power before the objective lens, $T_0$=0.83 the optical transmittance of the objective lens at 415 nm\textemdash the pump wavelength\textemdash and $R_0$=0.444 the optical reflection coefficient of Ti at 415 nm),  $t$ the bilayer thickness (with $t_\text{Ti}=660$, $t_\text{Si$_3$N$_4$}=1830$~nm), $\kappa=27.8$~W/m.K the thermal conductivity estimated by weighting the values for each layer by their thickness ($\kappa_\text{Ti}=21.9$, $\kappa_\text{Si$_3$N$_4$}=30$~W/m.K), $a=5.64$~mm the plate radius (the circular plate being chosen to have the same area as the square sample plate surface $10\times10$~mm$^2$) and $w=4.2$~\textmu m the $1/e^2$ intensity radius of the pump beam. Reflection coefficients and thermal conductivities are taken from Ref.~\citenum{lide10livre}}. The Lamb waves thermoelastically generated with our pump are calculated to have an amplitude of $\sim$10~pm. To determine the dispersion relation experiments are also carried out using a pump spot in the form of a line of 1/$e^2$ intensity full-width 1.5~\textmu m and a length of 5~\textmu m.

The other part of the beam, at $\lambda=830$~nm, is used for the probe. It is focused  to a 2.8~\textmu m radius (at $1/e^2$ intensity) for both types of pump beam focusing
with the same objective lens as used for the pump and with a $\sim$0.03~nJ pulse energy. When required, it is intensity modulated with another AOM at the frequency $f_s$. The modulation frequency is set to allow heterodyne detection within the 3~MHz photodetector bandwidth ($|f_p-f_s|\leqslant3$~MHz). A Sagnac interferometer is incorporated \cite{tachizaki06}, producing two probe pulses separated by a time interval of $t<200$~ps. A delay line and a 2D spatial scanning system making use of a two-axis displacement stage with a lens pair are mounted in the path of the probe beam to access the spatiotemporal evolution of the acoustic field at the sample surface at 20~Hz bandwidth. The out-of-plane surface particle velocity modulates the optical phase, which is converted to an intensity modulation that is detected with a lock-in amplifier. The arbitrary-frequency technique allows one to access frequencies $n f_\text{rep}+m f_p$, $n$ an integer and $m=\pm1$, by recording both in-phase and  quadrature components~\cite{matsuda15, kaneko14, mezil15a}.\\

\noindent\textbf{Data analysis ---}
The setup used here makes use of the following: double modulation
(i.e. intensity modulation of both pump and probe beams), a delay set
in the probe-beam line, and a probe modulation upstream of the
delay. The arrangement is described in detail in
Ref.~\citenum{matsuda15}.  Because of the pump intensity modulation,
the excited frequencies are $n f_\text{rep}+m f_p$, where $m=+1$ for
the upper sideband and $-1$ for the lower sideband.  The generated
acoustic field can be expressed as a superposition of these frequency
components as
\begin{equation}
S(t)=\sum_{n,m}A_{n,m}\cos[-(n\omega_\text{rep}+m\omega_p)t+\phi_{n,m}],
\label{eq:FT}
\end{equation}
where $A_{n,m}$ and $\phi_{n,m}$ are the position dependent amplitude
and phase of a vibrational mode
$(n,m)$ of frequency $n f_\text{rep}+m f_p$.
In the summation of Eq.~\ref{eq:FT}, the case $(n,m)=(0,-1)$ is not included.
From the detection system (i.e. a photodetector connected to a lock-in amplifier), we access the in-phase ($X$) and  quadrature ($Y$) components of the lock-in output, which are proportional to the acoustic out-of-plane surface particle velocity\footnote{We assume $X\propto\int u(t)\cos\omega_\text{ref}\, tdt$ and $Y\propto\int u(t)\sin\omega_\text{ref}\, t dt$ where $u(t)$ is the signal from the photodetector and $\omega_\text{ref}=|\omega_s-\omega_p|$.}. Both are functions of delay time $\tau$ and probe spot position. The complex signal $Z\equiv X+iY\text{sgn}(\omega_p-\omega_s)$ can be rewritten as 
\begin{equation}
Z(\tau)=\sum_{n,m} A_{n,m}e^{im[-(n\omega_\text{rep}+m\omega_s)\tau+\phi_{n,m}]}.
\label{eq:Z}
\end{equation}
In the summation of Eq.~\ref{eq:Z}, again the case $(n,m)=(0,-1)$ is not included.
The amplitude and the phase are found from a Fourier transform (FT):
\begin{equation}
A_{n,m}e^{im\phi_{n,m}}=\frac{1}{T}\int^T_0Z(\tau)e^{i(mn\omega_\text{rep}+\omega_s)\tau}d\tau.
\end{equation}
It is then possible to analyze the data at an angular frequency $\omega$ as a
real-valued term involved in Eq.~\ref{eq:FT}, i.e., giving a spatial pattern of
the vibration at $\omega=2\pi(n f_\text{rep}+m f_p)$. Modifying the frequency $f_p$ thus allows arbitrary acoustic frequency control.

In the first experiment, the pump frequency $f_p$ is increased from 0.1 up to 4~MHz by steps of 0.1~MHz. This provides data at frequencies $n f_\text{rep}+m f_p$. In the spectrum displayed in Fig.~\ref{fig:evolpeak}, one can see the amplitude evolution associated with $n=21$ and $m=+1$ for all these pump frequencies. The maximum amplitude is obtained for $f_p=2.0$~MHz. In the following experiments  the pump frequency is maintained at $f_p=2.0$~MHz to optimize the ZGV Lamb mode generation. The resulting measured acoustic frequencies are $n f_\text{rep}\pm 2.0$~MHz. In Fig.~\ref{fig:fouriersignal} the analysis covers all the scanned frequencies, but for each $n$ only the maximum amplitude is displayed for clarity (e.g., for $n=21$ only the point associated with $f_p=2.0$~MHz is displayed).

\section*{References}
\vspace{-5mm}
%

\vspace{-2mm}
\section*{Acknowledgements}
\vspace{-5mm}
\noindent The authors are grateful to Claire Prada and Alex Maznev for fruitful discussions. S. Mezil carried out this work as an International Research Fellow of the Japanese Society for the Promotion of Science (JSPS).  

\vspace{-2mm}
\section*{Author contributions}
\vspace{-5mm}
\noindent O.B.W. and S.M. proposed the research goals and supervised the project. Q.X. performed experiments with the help of S.M. and M.T. The theoretical model and numerical model was developed by S.M. and J.L. Data was analyzed by Q.X., S.M. and P.H.O. Theoretical support was provided by S.M., O.M., Z.S. and O.B.W. All authors helped prepare, critically review and revise the manuscript.  

\vspace{-2mm}
\section*{Additional information}
\vspace{-5mm}
\noindent\textbf{Supplementary Information} accompanies this paper.\\
\noindent\textbf{Multimedia animations} are available at \href{https://www.dropbox.com/s/uxu1avhfhpnchez/anim.gif?dl=0}{this link}.\\
\noindent\textbf{Competing interests:} The authors declare no competing interests.

 \end{document}


\vspace*{\baselineskip}
\vspace*{\baselineskip}

\title{Supplementary Information \\
 Imaging GHz Zero-Group-Velocity Lamb Waves
 \vspace*{\baselineskip}}


\author{Qingnan Xie}
\affiliation{School of Science, Nanjing University of Science and Technology, Nanjing 210094, People's Republic of China}
\author{Sylvain Mezil$^*$}
\author{Paul H. Otsuka}
\author{Motonobu Tomoda}
\affiliation{Division of Applied Physics, Graduate School of Engineering, Hokkaido University, Sapporo 060-8628, Japan}
\author{J\'er\^ome Laurent}
\affiliation{Institut Langevin, ESPCI ParisTech, CNRS, 1 rue Jussieu, 75238 Paris Cedex 05, France}
\author{Osamu Matsuda}
\affiliation{Division of Applied Physics, Graduate School of Engineering, Hokkaido University, Sapporo 060-8628, Japan}
\author{Zhonghua Shen}
\affiliation{School of Science, Nanjing University of Science and Technology, Nanjing 210094, People's Republic of China}
\author{Oliver B. Wright}
\affiliation{Division of Applied Physics, Graduate School of Engineering, Hokkaido University, Sapporo 060-8628, Japan}


\begin{abstract}
\vspace*{1cm}
We describe the model used to predict the dispersion curves and displacements for a bilayer system with perfect coupling between the layers. We also present pulse-echo measurements used to derive the layer thicknesses. We apply the theoretical model to predict the ZGV Lamb mode frequencies, their wavenumbers and their out-of-plane and in-plane displacements. 
\vspace*{\baselineskip}

\end{abstract}

\maketitle

 \section*{}
 \newpage
 \section*{}
 \newpage

\section*{Theoretical model}

\begin{figure}[h]
\centering
\includegraphics[scale=1]{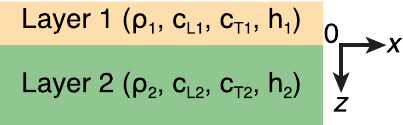}
\caption{\textbf{Geometry of the bilayer model.}} 
\label{fig:bilayer}
\end{figure}

The geometry is shown in Fig.~\ref{fig:bilayer}. We assume that the two layers are isotropic, homogeneous and infinite, with mass density $\rho_i$, longitudinal and transverse velocities $c_{Li}$ and $c_{Ti}$, and thicknesses $h_i$, where $i=1,2$ indicates the layer number. The coupling between the layers is taken to be perfect, i.e., by assuming continuity of the displacement and stress components at the interface ($z=0$). The $\omega$ (angular frequency) -- $k$ (wavenumber) relation is solved using the scalar potential $\phi$ and the vector potential $\psi$, where the latter is reduced to a scalar as the problem is two-dimensional. The tangential and normal displacements are derived from these potentials as follows:
\begin{equation}
u_{x}=\frac{\partial\phi}{\partial x}	-\frac{\partial\psi}{\partial z}\text{, }\quad u_{z}=\frac{\partial\phi}{\partial z}+\frac{\partial\psi}{\partial x},
\label{eq:ux_uz}
\end{equation}
and the stresses are given by
\begin{eqnarray}
\sigma_{xz}&=&\mu\left(\frac{2\partial^2\phi}{\partial x\partial z}+\frac{\partial^2\psi}{\partial x^2}-\frac{\partial^2\psi}{\partial z^2}\right),\\
\sigma_{zz}&=&\lambda\left(\frac{\partial^2\phi}{\partial x^2}+\frac{\partial^2\phi}{\partial z^2}\right)+2\mu\left(\frac{\partial^2\phi}{\partial z^2}+\frac{\partial^2\psi}{\partial x\partial z}\right),
 \end{eqnarray}
where $\lambda$, $\mu$ are the Lam\'e coefficients \cite{achenbach73ou84livre}.

The potentials in the layers can be expressed as

\begin{equation}
\begin{cases}
\phi_1=\left[A_{1L}\cos(p_1 z)+B_{1L}\sin(p_1 z)\right]e^{\jmath\left(k x-\omega t\right)}\\
\psi_1=\left[A_{1T}\cos(q_1 z)+B_{1T}\sin(q_1 z)\right]e^{\jmath\left(k x-\omega t\right)}\\
\phi_2=\left[A_{2L}\cos(p_2 z)+B_{2L}\sin(p_2 z)\right]e^{\jmath\left(k x-\omega t\right)}\\
\psi_2=\left[A_{2T}\cos(q_2 z)+B_{2T}\sin(q_2 z)\right]e^{\jmath\left(k x-\omega t\right)},
\label{eq:psiphi}
\end{cases}
\end{equation}
where $p$ and $q$ are the $z$-components of the longitudinal and transverse wave vectors, respectively. The wavenumbers $k_{Li}=\omega/c_{Li}$ and $k_{Ti}=\omega/c_{Ti}$ satisfy  dispersion relations for bulk waves  ${k_{Li}}^2={k}^2+{p_i}^2$ and ${k_{Ti}}^2={k}^2+{q_i}^2$. $A_{iL}$, $B_{iL}$ are the amplitudes of longitudinal components and $A_{iT}$, $B_{iT}$ are the amplitudes of shear components.

At the free boundaries ($z=-h_1$ and $h_2$),  the stresses normal to the surface ($\sigma_{xz}$ and $\sigma_{zz}$) vanish, whereas at the  interface ($z=0$),  the continuity of displacement and stresses is applied. It  follows that
\begin{equation}
\begin{cases}
\sigma_{zz1}=\sigma_{xz1}=0 & \text{for $z=-h_1$},\\
\sigma_{zz1}=\sigma_{zz2} &  \text{for $z=0$},\\
\sigma_{xz1}=\sigma_{xz2} & \text{for $z=0$},\\
u_{x1}=u_{x2} & \text{for $z=0$},\\
u_{z1}=u_{z2} & \text{for $z=0$},\\
\sigma_{zz2}=\sigma_{xz2}=0 & \text{for $z=h_2$}.
\label{eq:condlim}
\end{cases}
\end{equation}

From Eqs.~(\ref{eq:ux_uz}--\ref{eq:condlim}), the problem can be rewritten in matrix form, $\mathbf{M}\cdot\mathbf{U}=[0]$:
\begin{widetext}
\footnotesize
\centering
\begin{displaymath}
\left[ \begin{array}{cccc}
2 \jmath k p_1 \sin[p_1 h_1]	& 		2 \jmath k  p_1 \cos[p_1 h_1] 			&	 ({k_{t1}}^2-2 k^2) \cos[q_1 h_1] 		&		 -({k_{t1}}^2-2 k^2) \text{sin}[q_1 h_1]  \\
-({k_{t1}}^2-2 k^2) \cos[p_1 h_1] & 		({k_{t1}}^2-2 k^2) \text{sin}[p_1 h_1] 		&	 2 \jmath k q_1 \sin[q_1 h_1] 			&		 2 \jmath k q_1 \cos[q_1 h_1] \\
0 				                 & 		2 \jmath k \mu_1 p_1 				& 	({k_{t1}}^2-2 k^2) \mu_1 				& 		0\\
-({k_{t1}}^2-2 k^2) \mu_1 		& 		0 								&	 0 					                    	& 		2 \jmath k \mu_1 q_1\\
\jmath k 				        & 		0 								& 	0 								& 		-q_1 \\
 0 						& 		p_1 								& 	\jmath k 						         & 		0 \\
 0 						& 		0 								& 	0 								& 		0 \\
 0 						& 		0 								& 	0 								& 		0
\end{array} \right.
\end{displaymath}
\centering
\begin{multline}
\left. \begin{array}{cccc}
0 						& 		0 								& 	0 								& 		0\\
0						& 		0								&	0								&		0\\
0						& 		-2\jmath k \mu_2 p_2				&	-({k_{t2}}^2-2k^2) \mu_2				&		0\\
({k_{t2}}^2-2k^2)\mu_2		&		0								& 	0								&		-2\jmath k \mu_2 q_2\\
-\jmath k					&		0								&	0								&		q_2\\
0						&		-p_2								&	-\jmath k							&		0\\
-2\jmath k p_2 \sin[p_2 h_2]	&		2\jmath k p_2 \cos[p_2h_2]			&	({k_{t2}}^2-2k^2)\cos[q_2h_2]			&		({k_{t2}}^2-2k^2)\text{sin}[q_2h_2]\\
-({k_{t2}}^2-2k^2)\cos[p_2h_2] 	&		-({k_{t2}}^2-2k^2)\text{sin}[p_2h_2]		&	-2\jmath k q_2 \sin[q_2 h_2]			&		2 \jmath k q_2 \cos[q_2 h_2]
\end{array} \right]\cdot
\begin{bmatrix}
A_{1L} \\ B_{1L} \\ A_{1T} \\ B_{1T} \\  A_{2L} \\ B_{2L} \\ A_{2T} \\ B_{2T}
\end{bmatrix}=
\begin{pmatrix}
0 \\ 0 \\ 0 \\ 0 \\ 0 \\0\\0\\0
\end{pmatrix}.
\label{eq:matriceS}
\end{multline}
\end{widetext}

\begin{figure*}
\centering
\includegraphics[width=\textwidth]{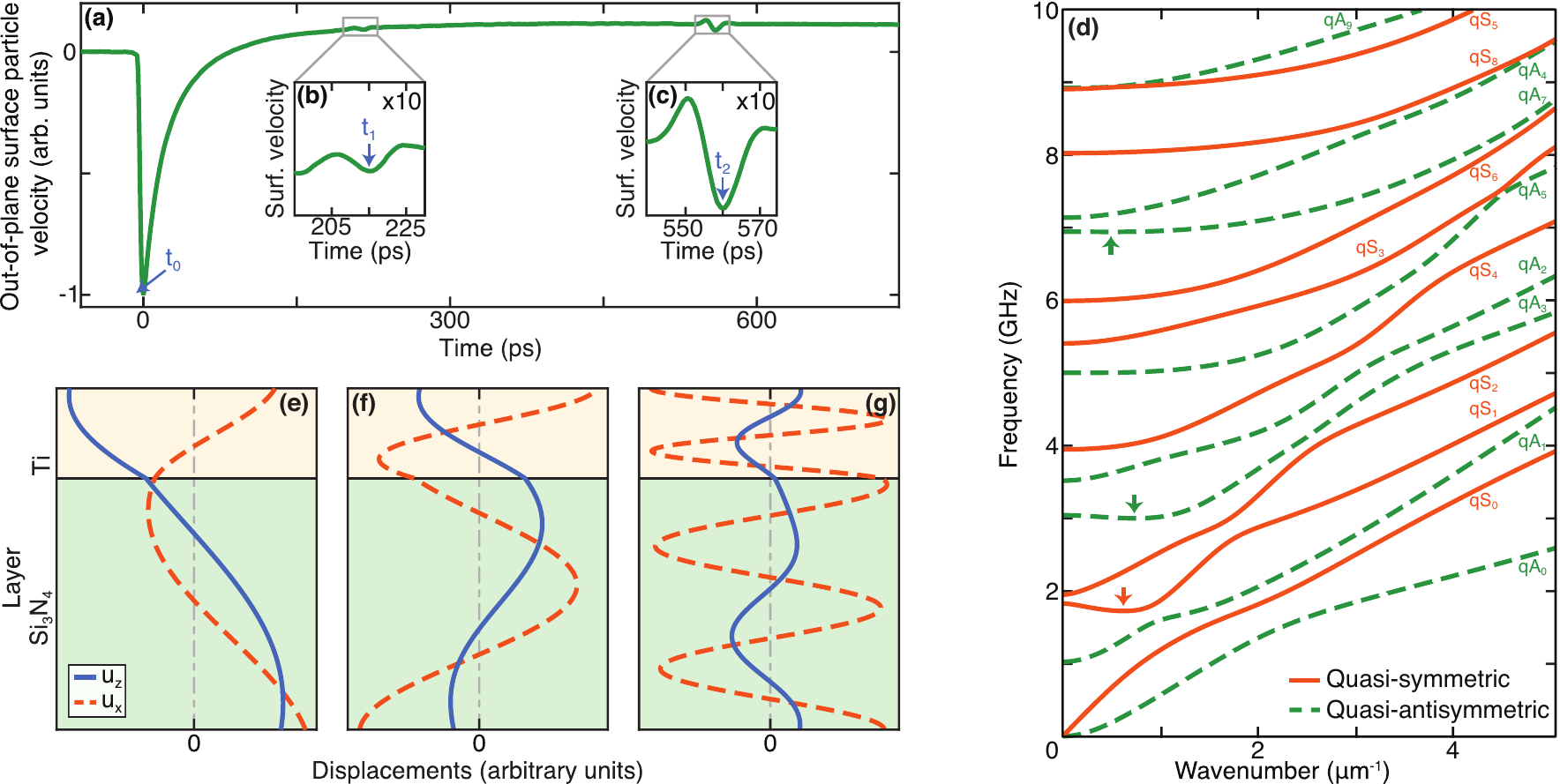}
\subfigure{\label{fig:thickness}}
\subfigure{\label{fig:thicknesszoom1}}
\subfigure{\label{fig:thicknesszoom2}}
\subfigure{\label{fig:disptheo}}
\subfigure{\label{fig:displac169}}
\subfigure{\label{fig:displac300}}
\subfigure{\label{fig:displac694}}
\caption{\textbf{Pulse-echo measurements and ZGV mode displacements.} (a)~Surface particle velocity temporal variation, showing acoustic echoes. Zoom-in on the echoes from (b)~the interface and (c)~the rear surface of the sample. (d)~Calculated dispersion curves of the bilayer system.  (e--g)~Calculated normal (solid line) and tangential (dashed line) displacements in the Ti/\SiN bilayer for the first three ZGV Lamb modes (a)~at $f_1^\text{th}=1.7248$~GHz and $k_1^\text{th}=0.620$~\textmu m$^{-1}$ (b)~at $f_2^\text{th}=3.0014$~GHz and $k_2^{\text{th}}=0.732$~\textmu m$^{-1}$ and (c)~at $f_3^{\text{th}}=6.9476$~GHz and $k_3^{\text{th}}=0.492$~\textmu m$^{-1}$.}
\end{figure*}

Non-trivial solutions are found when the determinant of the $8\times8$ matrix $\mathbf{M}$  vanishes, i.e.,  det$(\mathbf{M})=0$. In order to avoid (unwanted) bulk waves propagating at velocities $c_{Li}$ ($p_i=0$) and $c_{Ti}$ ($q_i=0$), the terms $p_1$, $q_1$, $p_2$ and $q_2$ can be factorized in the $2^\text{nd}$, $4^\text{th}$ $6^\text{th}$ and $8^\text{th}$ rows, respectively. 
The dispersion curves of the bilayer structure is then estimated by determining the zeros of the secular equation. As the structure is spatially asymmetric, modes cannot be classified exactly as symmetric and antisymmetric. For a given mode, the group velocity is extracted using $v_g=\partial\omega/\partial k$. A solution  $(\omega, k)$ is identified as a ZGV mode if $v_g=0$ with $k\neq0$. Furthermore, normal and tangential displacements---$u_z$ and $u_x$, respectively---can be estimated from the dispersion curves. For a solution $(\omega,k)$, the equations representing the boundary conditions can be be solved once a component common to $\bf{U}$ is fixed (e.g., $A_{1L} = 1$). This gives access to the relative displacements $u_{x,z}$.

\section*{\label{sec:param}Sample and experimental parameters}

\noindent The sample consists of a silicon-nitride membrane provided by NTT Advanced Technology Corporation (MEM-N0302) with a nominal thickness of 
$2\pm 0.2$~\textmu m. 
It is mostly composed of \SiN, but is not a pure crystal (the composition ratio  Si:N is between 3:4 and 1:1). Nevertheless, it is hereafter denoted as \SiN.  The membrane is supported on its edges by a Si frame, providing a $3\times3$~mm$^2$ area with free surfaces, necessary to generate ZGV Lamb modes. The membrane is coated with a $\sim$500~nm sputtered polycrystalline titanium film. 
To calculate the dispersion curves, the elastic constants and density are taken from Ref.~\citenum{briggs92livre}: $c_{L_{Ti}}\!=\!6130$~m.s$^{-1}$, $c_{T_{Ti}}\!=\!3182$~m.s$^{-1}$, $\rho_{_{Ti}}\!=\!4508$~kg.m$^{-3}$ for titanium and $c_{L_{Si_3N_4}}\!=\!10607$~m.s$^{-1}$, $c_{T_{Si_3N_4}}\!=\!6204$~m.s$^{-1}$, $\rho_{_{Si_3N_4}}\!=\!3185$~kg.m$^{-3}$ for silicon nitride. 
In order to accurately determine the thicknesses, an experiment measuring the surface particle velocity in the time domain is carried out using an interferometric pulse-echo method with focused pulsed-laser beams  ($\sim$1.5~\textmu m $1/e^2$ diameter) and picosecond time resolution. The pump beam is modulated at $f_p=1$~MHz, and we monitor the in-phase output of the lock-in amplifier. The result is shown in Fig.~\ref{fig:thickness}. The first minimum in the variation at $t_0=0$ is related to the temperature rise and deformation caused by the laser pulse. The echo at $t_1$ corresponds to the acoustic pulse reflected from the \SiN/Ti interface, whereas the second echo at $t_2$ corresponds to the acoustic pulse reflected from the rear surface of the membrane. The  weak reflection from the interface (at $t_1$) indicates good adhesion (as our model assumes). The corresponding time intervals are $\Delta t_1= 215\pm1$~ps and $\Delta t_2 =560\pm1$~ps, allowing us to evaluate the thicknesses  of $659\pm 3$ and $1830\pm 10$~nm for the Ti and the \SiN layers, respectively, from the known $c_{L}$ values\footnote{Errors correspond to those arising from the time resolution of the apparatus.}. For \SiN the thickness agrees within the 10\% uncertainty given by the supplier. 
\begin{table}[t]
\begin{minipage}[c]{.88\columnwidth}
\begin{ruledtabular}
\centering
\caption{First three ZGV Lamb mode frequencies $f$ and wavenumbers $k$ for the Ti/\SiN bilayer.}
\label{tab:values}
\begin{tabular}{cccc}
& Mode &  $f$ (GHz) & $k$ (\textmu m$^{-1}$)  \\
 \hline
1 & $qS_1$ & 1.7248 		      & 0.620       \\  
2 & $qA_3$ &  3.0014            & 0.732          \\      
3 & $qA_7$ &  6.9476            & 0.492                  
\end{tabular}
\end{ruledtabular}
\end{minipage}
\end{table}

The corresponding predicted dispersion curves are shown in Fig.~\ref{fig:disptheo}. The mode classification follows the one suggested by Mindlin~\cite{mindlin2006introduction}, where the integers correspond to the number of antinodes of the mechanical displacement. This integer can be negative in case of negative group velocity. The `q' denomination relates to the term quasi- in the appellations quasi-symmetric  and quasi-antisymmetric, related to the sample spatial asymmetry.  For the solid modes (i.e., for $k=0$~\textmu m$^{-1}$),  $qA_{2n}$, $qS_{2n+1}$ have an out-of-plane displacement whereas $qA_{2n+1}$, $qS_{2n}$ have an in-plane displacement. Therefore, the former are more likely to be observed in our experiments. Three ZGV Lamb modes are predicted below 10~GHz. They are then referred as $qS_1$, $qA_3$ and $qA_7$, and are  labeled $1$, $2$, $3$, respectively,  for simplicity (see Table~\ref{tab:values}).
Their frequencies and associated wavenumbers are displayed in Table~\ref{tab:values}. 
With the  arbitrary-frequency method (see Methods in the Main text), these frequencies are accessible by modulating the pump beam at the frequency $f_p=36.8$, $f_p=27.3$, and $f_p=34.9$~MHz, for the first, second and third ZGV Lamb modes, respectively. 

We also present the normal and tangential displacements of these three ZGV modes in Fig.~\ref{fig:displac169}--\subref{fig:displac694}. 
At the top free surface, i.e., where the excitation and detection occur, the tangential displacement is significant for the three modes.  Conversely, the normal displacement is different for these modes: it is predominant for the lowest ZGV mode at $f_1^\text{th}=1.7248$~GHz (Fig.~\ref{fig:displac169}), still significant for the second one at $f_2^\text{th}=3.0014$~GHz (Fig.~\ref{fig:displac300}) and relatively weak for the third one at $f_3^\text{th}=6.9476$~GHz (Fig.~\ref{fig:displac694}).

Finally, the pump beam radius should be carefully chosen to enhance ZGV Lamb mode generation. For a single isotropic plate, Bruno \textit{et al.} demonstrated that, for a Gaussian beam, the optimum response is 
reached when the $1/e^2$ radius is $2\sqrt{2}/k$~\cite{bruno16}. Extending this result for our bilayer system leads to an ideal pump radius of $\sim$4.6~\textmu m for the first ZGV mode. In our set-up, detection sensitivity is inversely proportional to the probe beam radius. As both pump and probe beams are focused with the same objective lens (see Fig.~1(a) in the main text), 
 it is difficult to achieve the ideal case. A good compromise is found with the pump and probe $1/e^2$  radii, measured by knife-edge technique, set to be 4.2 and 2.8~\textmu m, respectively. 
This optimises the generation of propagating modes with wavenumber $k=0.67$~\textmu m$^{-1}$, but modes with $0.3\leqslant k\leqslant 1.4$ should also be generated. For the particular case of solid modes ($k=0$), we can expect such modes generated if the radius is larger than 2.1~\textmu m. These limits are estimated by extrapolating the numerical results found by Balogun \textit{et al.} in a 50~\textmu m thick aluminum plate~\cite{balogun07}. Generation is considered effective, as in this case, when the normalised amplitude (associated to the wavenumber and radius) is greater than half of its maximum value. 

\bibliography{bibliographiecomplete}{}